\documentclass[aps,twocolumn,prb,reprint,amsmath,amssymb,showpacs,superscriptaddress]{revtex4-1}

\usepackage{graphicx}
\usepackage{dcolumn}
\usepackage{bm}
\usepackage{amsmath}
\usepackage{color}
\usepackage{mathrsfs}
\usepackage{float}
\usepackage{dsfont}
\usepackage{txfonts}
\usepackage{wasysym}
\usepackage{ifsym}

\usepackage{amsmath}
\newcommand{\Luu}{\Lambda^{\uparrow \uparrow}}
\newcommand{\Lud}{\Lambda^{\uparrow \downarrow}}
\newcommand{\Ldu}{\Lambda^{\downarrow \uparrow}}
\newcommand{\Ldd}{\Lambda^{\downarrow \downarrow}}

\newcommand{\Lsp}{\Lambda^{\sigma \sigma'}}
\newcommand{\Lps}{\Lambda^{\sigma' \sigma}}

\newcommand{\Csp}{\chi^{\sigma  \sigma'}}

\newcommand{\dVsig}{\dot V_\sigma}
\newcommand{\dVpig}{\dot V_{\sigma'}}

\newcommand{\ndw}{n_\downarrow}
\newcommand{\nup}{n_\uparrow}
\newcommand{\nsig}{n_{\sigma }}
\newcommand{\npig}{n_{\sigma'}}

\newcommand{\nfsig}{n_{f \sigma }}

\renewcommand{\H}{\hat H}

\usepackage{graphicx}
\usepackage{braket}

\newcommand{\trace}[1] {\text{Tr} \left[ #1 \right]}
\newcommand{\expval}[1] {\left< #1 \right>}
\newcommand{\comm}[2] {\left[ #1, #2 \right]}

\newcommand{\ketbra}[2]{\ket{#1}\bra{#2}}

\newcommand{\intInf}{\int_{-\infty}^{\infty}}
\newcommand{\intKubo}{i\int_{-\infty}^{\infty} dt' \Theta(t-t')}

\renewcommand{\vec}[1]{\mathbf{#1}}

\newcommand{\fresp}[2]{-i\Theta(t-t')(\vec r', t')}

\newcommand{\im}[1]{\text{Im}\left \lbrace #1 \right \rbrace}

\makeatletter

\newcommand{\Rmnum}[1]{\expandafter\@slowromancap\romannumeral #1@}
\makeatother
\begin{document}

\title{Work exchange, geometric magnetization  and fluctuation-dissipation relations in a quantum dot under adiabatic magnetoelectric driving}
\author{Pablo Terr\'en Alonso}
\affiliation{International Center for Advanced Studies, Escuela de Ciencia y Tecnolog\'{\i}a, Universidad Nacional de San Mart\'{\i}n-UNSAM, Av 25 de Mayo y Francia, 1650 Buenos Aires, Argentina} 

\author{Javier Romero}
\affiliation{International Center for Advanced Studies, Escuela de Ciencia y Tecnolog\'{\i}a, Universidad Nacional de San Mart\'{\i}n-UNSAM, Av 25 de Mayo y Francia, 1650 Buenos Aires, Argentina}

\author{Liliana Arrachea}
\affiliation{International Center for Advanced Studies, Escuela de Ciencia y Tecnolog\'{\i}a, Universidad Nacional de San Mart\'{\i}n-UNSAM, Av 25 de Mayo y Francia, 1650 Buenos Aires, Argentina} 

\begin{abstract}
We study the adiabatic dynamics of the charge, spin and energy of a quantum dot with a Coulomb interaction under two-parameter driving, associated to  time-dependent gate voltage  and 
magnetic field. The quantum dot is coupled to a single reservoir at temperature $T=0$ and the dynamical Onsager matrix is fully symmetric, hence, the net energy dynamics  is fully dissipative. However, in the presence of many-body interactions, other interesting mechanisms 
take place, like the net exchange of work between the two types of forces and the non-equilibrium accumulation of charge with different spin orientations. The latter has  a 
geometric nature.  The dissipation takes place in
the form of an instantaneous Joule law with the universal resistance $R_0=h/2e^2$. 
We show the relation between this Joule law and instantaneous fluctuation-dissipation relations. The latter lead to generalized  Korringa-Shiba relations, valid in the Fermi-liquid regime. 
\end{abstract}

\date{\today}

\maketitle

\section{Introduction}
Understanding the charge  dynamics in driven quantum dots  motivates intense research activity for some years now.
Several developments took place after B\"uttiker and coworkers discussed the dynamics of the quantum RC circuit composed by a quantum capacitor  in series with a resistor, which operates under the driving by 
an AC voltage. \cite{btp1,btp2} Not only the charge but also the energy dynamics in this system deserves attention. In fact, this is the minimal electronic system to address fundamental questions on the energy transport
and heat production in the quantum regime, which is a subject under active investigation across the areas of statistical mechanics, condensed matter, cold atoms and 
quantum information.

The minimal RC quantum circuit can be realized in a quantum dot (QD). The charge dynamics of this circuit under AC driving  has been the subject of several experimental \cite{qcap1,qcap2,qcap3,qcap4} and theoretical \cite{ringel,mora,hama,rosa2,mich1,rossello,dut,janine,mich2,mich3,nonlin,misha1,misha2,misha3,misha4,re1,adia-nice,entro,keel,piet,david,rome}
studies
 in the linear and non-linear regimes.
The capacitive element is, precisely, the QD, while the resistive element is the contact to a fermionic reservoir. The dynamics of the charge
in linear response is ruled by the capacitance of the quantum dot and the universal B\"uttiker resistance $R_0=h/2e^2$. 
When the AC driving potential is applied at the QD, a net amount of energy is dissipated in the form of heat. 
Remarkably, several recent works indicate that for adiabatic driving -- when the characteristic time of the driving is much larger than the one of the electrons in the QD--  and the reservoir is at temperature $T=0$, such heat generation
follows an instantaneous 
Joule law (IJL)  with the universal resistance $R_0$ per conducting channel. 
This result holds for a non-interacting QD, \cite{re1,adia-nice,entro} as well as for a quantum dot with Coulomb interactions, \cite{rome} and extends to more complex configurations
 containing normal and superconducting leads. \cite{liro} In the case of  QD with many-body interactions and single-parameter driving, the proof of the universal IJL relies on an identity known as the Korringa-Shiba relation, \cite{ks} 
 derived for Fermi liquids. After Ref. \onlinecite{re1}, several studies also focused on the distribution of the heat production along the different pieces of the device and the role of the energy reactance. \cite{espo1, gnomo, adia-nice,entro,espo2, reac,tran}

  \begin{figure}[H]
  \includegraphics[width=\columnwidth]{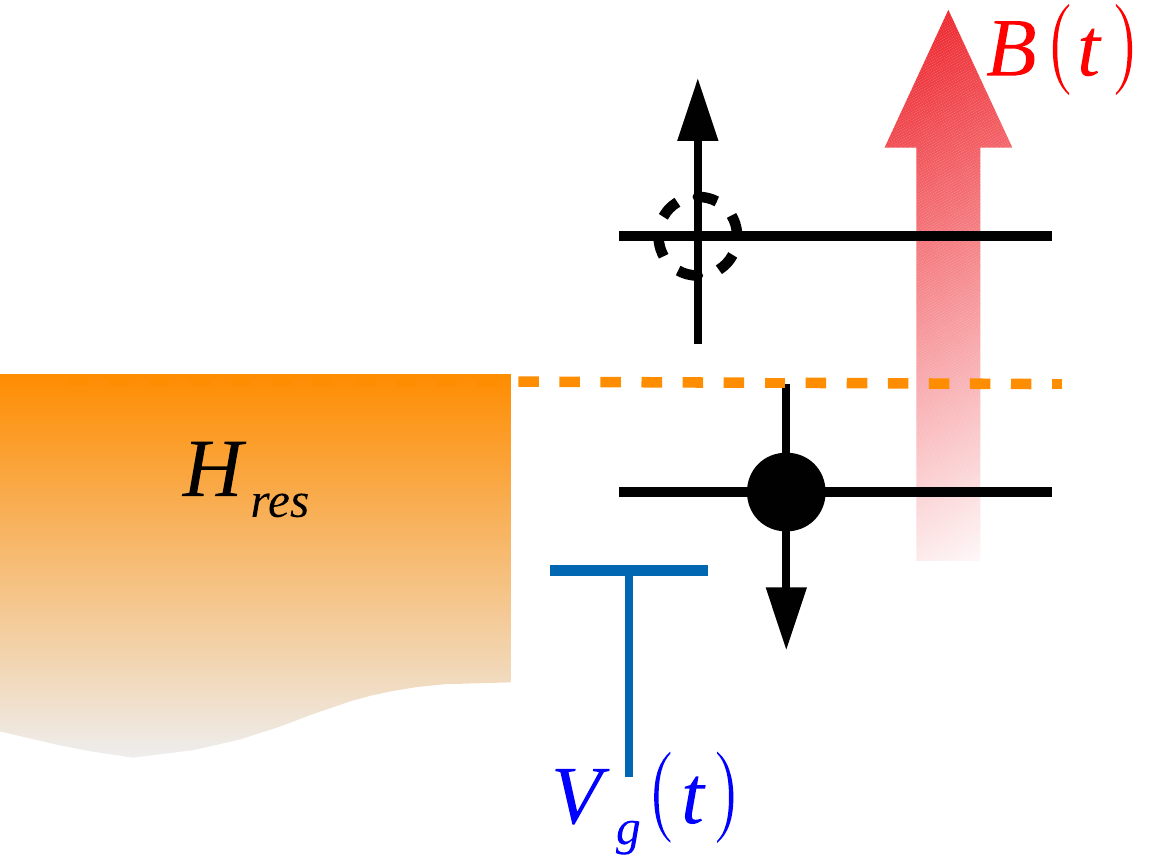}
  \caption{(Color online) Sketch of the setup. A quantum dot  connected to a single electronic reservoir  at $T=0$ is driven by means of a gate voltage $V_g(t)$ and a magnetic field $B(t)$. We model the device by an Anderson Hamiltonian with a single level and Coulomb interaction $U$. }
  \label{fig0}
\end{figure}
 
 So far, all studies of the RC quantum circuit focus on driving protocols with a single time-dependent parameter, corresponding the an AC gate voltage at the quantum dot.  Adiabatically driven systems with several time-dependent
 parameters enable more interesting effects. The most well known example is the quantum pumping of charge between two electron reservoirs in electron systems locally modulated by two time-dependent parameters. The net transfer of charge in this case can be characterized by a  geometric quantity akin to the Berry phase. \cite{avron,brouwer} Other geometrical aspects of adiabatically driven systems, like
 geometric magnetism,\cite{miche} adiabatic perturbation theory in 
closed systems,\cite{anatoli}  the thermodynamic metric in q-bit systems,\cite{marti} and rectification, \cite{geora} have been recently analyzed. Topological properties of the energy conversion have been so far considered
only beyond the adiabatic regime. \cite{gil}
 In the present work we study the RC circuit under two-parameter driving. In addition to the usual AC driving at the gate voltage, we include the effect of an AC magnetic field. A sketch 
 of the setup is presented in Fig. \ref{fig0}. We solve this problem with the adiabatic approach introduced in Ref. \onlinecite{adia}, which treats the non-equilibrium dynamics by expanding the evolution operator in powers of 
 the velocities of the time-dependent parameters. The coefficients of such expansion depend on  frozen equilibrium susceptibilities. In our case, the latter can be numerically exactly evaluated by recourse to numerical renormalization group (NRG),\cite{hewson,zitko}  as in Ref. \onlinecite{rome}.
 Alike to the case of  single-parameter driving, the  only dissipative element is the contact resistance. Hence, dissipation of energy follows an IJL with the universal resistance $R_0$ per spin channel. 
 We show that this fact, when analyzed from the perspective of  a fluctuation-dissipation relation,  implies a series of  relations between the static and  dynamical charge and spin susceptibilities, which constitute generalizations
 of the usual Korringa-Shiba law. We show that these relations are valid for both the interacting and non-interacting cases. The interesting  feature that 
  the two-parameter driving brings about is the non-equilibrium accumulation of a polarized charge in the QD. The latter can be characterized by a geometric quantity, akin to the pumped charge
 in quantum pumps. \cite{avron,brouwer} 
  We show that  many-body interactions are crucial for this effect to be realized in the present system. In fact the non-interacting limit effectively decouples into two single-parameter driven problems (one for each spin orientation).

  The work is organized as follows. We introduce the model in Section II. The theoretical approach is presented in Section III. Results are presented in Section IV. Section V contains summary and conclusions. Technical details are presented in Appendices A and B.

\section{Model}
We consider the
following Hamiltonian for the full setup of Fig. \ref{fig0},
\begin{equation}\label{model}
    \H(t) = \H_{dot}(t) + \H_{res} + \H_{coupling},
\end{equation}
which consists of an Anderson impurity model with driving at the impurity. 
The reservoir is modeled by 
free electrons  $\H_{res} = \sum_{k\sigma}\epsilon_{k}\hat c^\dagger_{k\sigma}\hat c_{k\sigma}$ and the coupling between the dot and the reservoir is $\H_{coupling}=w_c \sum_{k\sigma}\hat c^\dagger_{k\sigma}\hat d_{\sigma} + d^\dagger_{\sigma}\hat c_{k\sigma}$. The Hamiltonian for the QD reads
\begin{equation}
\label{eq:Hdot}
    \H_{dot}=\sum_\sigma V_\sigma(t) \hat n_\sigma + U(\hat n_\uparrow-1/2)(\hat n_\downarrow-1/2),
\end{equation}
where the effect of the driving is encoded in the parameters $V_\sigma(t) = \varepsilon_0+eV_g(t)+s_\sigma \mu_B B(t)$ being $s_\sigma=\pm1$ for $\sigma=\lbrace \uparrow, \downarrow \rbrace$.
The latter  includes the effect of the time-dependent gate voltage $V_g(t)$, which shifts the energy levels of the quantum dot as a function of time, and a time-dependent magnetic $B(t)$, which introduces a time-dependent
Zeeman splitting for the two spin orientations within the quantum dot.  The constants $e$ and $\mu_B$ are, respectively, the electron charge and the Bohr magneton. 
The Coulomb repulsion is denoted by $U$ and $\hat n_\sigma$ the dot number operator for electrons with spin $\sigma$. 

\section{Adiabatic response}
The adiabatic regime refers to changes in the  time-dependent parameters within a time scale which is much larger than the typical life-time of the electrons in the quantum dot. In the non-interacting case,  the latter is determined by the coupling $w_c$ and the density of states of the reservoir. In systems with many-body interactions, this quantity can be renormalized in a non-trivial way. Here, we follow Ref. \onlinecite{adia}, and characterize the adiabatic dynamics as the linear-response regime in
the "velocities" $\dot V_{\sigma}(t)$.

\subsection{Dynamics of charge, spin and work}
We denote $\nsig(t) = \expval{\hat n_\sigma}(t)$ the mean occupancy of the dot with electrons with spin $\sigma$ at time $t$. Following Refs.  \onlinecite{adia, rome}, the adiabatic dynamics for $ n_\sigma(t)$ is given by
\begin{equation}
\label{eq:nSigmaAdiabatico}
    n_{\sigma}(t)=n_{f\sigma}(t) + \sum_{\sigma'} \Lsp(t) \dot V_{\sigma'}(t)
\end{equation}
where $n_{f\sigma}(t) = \expval{\hat n_\sigma}_t$ is the \textit{frozen} occupancy of the dot, i.e. the occupancy evaluated with the equilibrium density matrix, corresponding to the Hamiltonian $\H_{f,t} = \H(t)$ frozen at
the time $t$.  In Appendix A we review the derivation of these results.
The adiabatic Onsager coefficients $\Lsp(t)$ can be computed from 
\begin{equation} \label{defl}
  \Lsp(t) = -\lim_{\omega \to 0} \frac{\im{\Csp_t(\omega)}}{\hbar \omega}.
\end{equation}
Here $\Csp_t(\omega)$ is the Fourier transform of the susceptibility $\Csp_t(t-t') = -i\Theta(t-t')\expval{\comm{\nsig(t)}{\npig(t')}}_t$, evaluated with the equilibrium frozen density matrix.

The local charge and magnetic moment at the QD are given by
\begin{eqnarray}
e n(t) & = & e \sum_{\sigma} n_{\sigma} (t), \nonumber \\
m(t) & = & n_{\uparrow} (t)- n_{\downarrow} (t).
\end{eqnarray}
Consequently, we can define the susceptibilities $\chi^c_t(t-t^{\prime}) = - i \theta(t-t^{\prime}) \langle \left[n(t), n(t^{\prime}) \right]\rangle$, $\chi^m_t (t-t^{\prime}) = - i \theta(t-t^{\prime}) \langle \left[  m(t), m(t^{\prime}) \right]\rangle$,
and $\chi^{cm}_t(t-t^{\prime}) = \chi^{mc}_t (t-t^{\prime})= - i \theta(t-t^{\prime}) \langle \left[ n(t), m(t^{\prime}) \right] \rangle$. From these susceptibilities, we can define the transport coefficients in a similar way as in Eq.
(\ref{defl}). These coefficients can be collected into a dynamical Onsager matrix 
\begin{equation}  \label{lamc}
\vec \Lambda_C(t) = 
\begin{pmatrix}
\Lambda_c(t) & \Lambda_{cm}(t) \\
\Lambda_{mc}(t) & \Lambda_m(t)
\end{pmatrix}.
\end{equation}
The latter  coefficients are related to the previously defined ones through
 \begin{eqnarray}
    \Lambda_c(t) &=& \Luu(t) +  \Ldu(t)+ \Lud(t)+ \Ldd(t), \nonumber \\
    \Lambda_{cm}(t) \;=\; \Lambda_{mc}(t)&=&\Luu(t) - \Ldd(t), \nonumber \\
    \Lambda_m(t) &=& \Luu(t)- \Ldu(t) - \Lud(t)+ \Ldd(t).
\end{eqnarray}

The current between the QD and the reservoir can be calculated from
\begin{equation}
\label{eq:iSigmaGeneral}
    I_\sigma(t) = e \dot n_\sigma(t),
\end{equation}
and reads
\begin{equation}
\label{eq:corrienteSigma}
 I_\sigma(t) = e\frac{d\nfsig(t)}{dt} + e\frac{d \left[ \sum_{\sigma'} \Lsp(t) \dVpig(t) \right]}{dt}.
\end{equation}

The AC power associated to the driving is $P(t)= \langle \partial \H /\partial t \rangle$. For the Hamiltonian (\ref{model}) it can be expressed as 
$P(t)=\sum_{\sigma} P_{\sigma}(t)$, with 
\begin{equation} \label{power}
P_{\sigma}(t)= n_{\sigma}(t) \dot{V}_{\sigma}(t).
\end{equation}
The  instantaneous occupancies $n_{\sigma} (t)$  play the role of conjugated forces to the driving potentials $V_{\sigma}(t)$. In fact, notice that the latter satisfy
$n_{\sigma} (t)= \langle {\partial H}/ {\partial V_{\sigma}} \rangle$.
From Eq. (\ref{power}) we see that, in the present problem, the power is determined by the dynamics of the charge with $\sigma$ polarization.
 In the adiabatic framework, Eq. (\ref{eq:nSigmaAdiabatico}) leads to 
\begin{equation}
\label{eq:pSigma}
 P_\sigma(t) = \nfsig(t)\dVsig(t) + \sum_{\sigma'} \Lsp(t) \dVpig(t) \dVsig(t).
\end{equation}

\subsection{Conservative and non-conservative geometric occupancies}
Let us focus on a cyclic driving protocol, where both driving fields depend on time with the same period $\tau= \Omega/2 \pi$, $V_g(t+ \tau)= V_g(t), \; B(t+\tau)=B(t)$, hence $V_{\sigma}(t+\tau)= V_{\sigma}(t)$. 

We can easily see that the first term $\propto \dVsig(t)$ in Eq. \eqref{eq:pSigma} is a conservative contribution, in the sense that it has a zero mean when averaged over one period. Explicitly,

\begin{equation}
P_{\rm cons} (t)= \sum_{\sigma} P_{\sigma, {\rm cons}}(t) = \sum_{\sigma} \nfsig(t)\dVsig(t).
\end{equation}
To see this, we define ${\bf V}(t)= \left(V_{\uparrow} (t), V_{\downarrow} (t)\right)$ and ${\bf n}_{f}(t)= \left( n_{f \uparrow} ({\bf V}), n_{f \downarrow} ({\bf V}) \right)$. 
These vectors are related 
to the mean value of the frozen Hamiltonian $H_{f} ({\bf V})  = \langle \H_{f,t} \rangle$ through ${\bf n}_{f}(t)= \partial_{\bf V} H_{f} ({\bf V})$.
In these expressions we have introduced a notation that highlights the fact that the time-dependence of $ \nfsig(t)$ is because of the time-dependent parameters $V_{\sigma}(t)$.  Then, we can express
the average over one period of the conservative component of the power as follows,
\begin{equation}
\overline{P_{\rm cons}}=   \frac{1}{\tau} \int_0^{\tau} dt \;  {\bf n}_f(t) \cdot \dot {\bf V}(t) = \oint_{\cal C} \partial_{\bf V} H_{f}({\bf V}) \cdot d {\bf V} =0,
\end{equation}
where $\oint_{\cal C}$ denotes the integral over a closed contour  in the parameter space ${\bf V}$.
Therefore ${\bf n}_f(t)$ is the conservative term of the force.

Similarly, the second term of Eq. (\ref{eq:pSigma}) contributes to the non-conservative component of the power. Introducing the definition of the matrix 
$\vec \Lambda (t)$, with matrix elements $\Lsp(t)$, the corresponding component of the force reads
\begin{equation} \label{n-noncons}
{\bf n}_{\rm non-cons}(t)=   \vec \Lambda ({\bf V})  \cdot  \dot{\bf V}(t).
\end{equation}
For a system with several parameters, this component  has a geometric character when averaged over one period,
\begin{equation}\label{non-cons}
\overline{\bf n}_{\rm non-cons}= \frac{1}{\tau} \int_0^{\tau}  dt \vec \Lambda ({\bf V})  \cdot  \dot{\bf V}(t) = \oint_{\cal C}  \vec \Lambda ({\bf V}) \cdot d{\bf V}.
\end{equation}
In the present problem, this implies a finite non-equilibrium accumulation of charge and a finite magnetization  
induced at the quantum dot, akin to the pumped charge in quantum pumps driven
by two or more parameters. \cite{avron,brouwer}

Quite generally, we can decompose the matrix  $ \vec \Lambda (t)= \vec \Lambda ^s(t) + \vec \Lambda ^a(t)$ into symmetric $ \vec \Lambda^s(t)$ and antisymmetric $ \vec \Lambda ^a(t)$
parts. Consequently, the force can be split into two components ${\bf n}_{\rm non-cons}(t)= {\bf n}^s_{\rm non-cons}(t)+ {\bf n}^a_{\rm non-cons}(t)$. The latter component is equivalent to a
Lorentz force associated to a geometric magnetic field, as discussed in Ref. \onlinecite{miche}.
 Only the symmetric component of the non-conservative force develops power, which  reads
\begin{equation}
P_{\rm non-cons} (t)= \dot{\bf V}^T(t) \cdot \vec \Lambda^s (t) \cdot  \dot{\bf V}(t),
\end{equation}

where the superscript $T$ stands for transposing the vector.

\subsection{Dissipation and work exchange}
We now turn to analyze the expected mechanisms in the dynamics of the energy. 
In the adiabatic regime,  we expect that the evolution satisfies the second principle of thermodynamics instantaneously,  the matrix $\Lambda^s (t)$ should be positive definite, implying that
its eigenvalues are $\lambda_m(t) \geq 0$. The instantaneously dissipated heat equals the total non-conservative power and reads 
\begin{equation}\label{eq:lam}
P_{\rm non-cons} (t) = \sum_m \lambda_m (t) \tilde{ \dot{{\bf V}}}(t)^2,
\end{equation}
where we have defined $\tilde{ \dot{{\bf V}} } (t)= {\bf U}  \dot{\bf V}(t)$, being ${\bf U}$ the unitary transformation that diagonalizes $ \vec \Lambda^s (t) $.

In a system with several driving parameters, it is possible to have exchange of work between the different induced forces, in addition to dissipation of energy. In fact, the non-conservative 
power developed by each force can be expressed as
\begin{eqnarray}\label{1}
P_{\rm non-cons, \sigma}(t) &=& \sum_{\sigma^{\prime}} \Lambda^{\sigma, \sigma^{\prime}} \dot{V}_{\sigma} (t)  \dot{V}_{\sigma^{\prime}} (t) ,\nonumber \\
& = &  P_{\rm diss, \sigma} (t) + s_{\sigma} P_{\rm exch} (t),
\end{eqnarray}
with $s_{\sigma}=\pm$. Hence, while it is satisfied
\begin{equation}\label{2}
\sum_{\sigma} P_{\rm non-cons, \sigma}(t) =  \sum_{\sigma} P_{\rm diss, \sigma} (t),
\end{equation}
 there is a finite amount of power $P_{\rm exch} (t)$, which 
can be exchanged between the two forces, associated to the charges with different spin orientations. Interestingly, this exchange may take place instantaneously, and also have  a net finite average over a period, 
\begin{equation}\label{netex}
\overline{P}_{\rm exch} = \frac{1}{\tau} \int_0^{\tau}P_{\rm exch} (t).
\end{equation}

\subsection{Relations between the transport coefficients}
Here we discuss the relations satisfied by the coefficients $\Lambda^{\sigma, \sigma^{\prime}}(t)$, which rule the adiabatic dynamics of the occupancy of the QD and the energy. These are 
Onsager relations, symmetry relations and fluctuation-dissipation relations. We analyze them separately.

\subsubsection{Onsager relations}
Since the susceptibilities entering the adiabatic dynamics are evaluated with the frozen Hamiltonian $H_{f,t}$ corresponding to the equilibrium condition at time $t$, they obey micro-reversibility, \cite{micro} which
leads to the following Onsager relation
\begin{equation}
\label{eq:simetriaLambdasObvia}
    \Lambda^{\sigma\sigma'}(V, B)=\Lambda^{\sigma'\sigma}(V, -B).
\end{equation}
The derivation is presented in Appendix A.

\subsubsection{Symmetry relations}
The Hamiltonian defined by Eqs. \eqref{model} and  \eqref{eq:Hdot}  
  is invariant under the transformation $\lbrace B,\downarrow,\uparrow \rbrace \rightarrow \lbrace -B,\uparrow,\downarrow\rbrace$.
This leads  to the following relations satisfied by the Onsager coefficients
\begin{equation}
\Lambda^{\sigma, \sigma^{\prime}}(V,B)= \Lambda^{\bar \sigma, \bar\sigma^{\prime}}(V,-B),
\end{equation}
where $\bar \downarrow = \uparrow$ and viceversa.
Notice that the cross susceptibilities, hence, the previous identity with $\sigma \neq \sigma^{\prime}$  vanish in the non-interacting limit. 

Combining this property with the Onsager relation Eq. \eqref{eq:simetriaLambdasObvia}, we find that the 
crossed coefficient satisfies
\begin{equation}
\label{eq:simetriaLambdasNoObvia}
    \Lambda^{\sigma\bar\sigma}(V, B)=\Lambda^{\bar\sigma\sigma}(V, B).
\end{equation}
 Notice that this relation implies that the matrix $ \vec \Lambda (t) $ is purely symmetric, i.e.  $ \vec \Lambda (t) \equiv \vec \Lambda^s (t) $.
 
\subsubsection{Fluctuation-dissipation relations at $T=0$}
Fluctuation-dissipation relations are usually explained by the energy balance. \cite{huang}
In the present problem, the only dissipation mechanism is the instantaneous Joule law due to the electron flow in the lead. This is characterized by a universal resistance $R_0$ per spin channel. Notice that in a
system with electron-electron interactions like the one considered here,
extra dissipation mechanisms might take place. However, the present model is known to be a Fermi liquid and such effects like inelastic-scattering are irrelevant within the low-energy regime, dominating the adiabatic dynamics. \cite{ks,mich2,rome}
We conclude that the instantaneous net dissipation reads
\begin{equation}\label{pjoulesigma}
P_{\rm diss, \sigma} (t) = P_{\rm Joule, \sigma} (t)= R_0 \left[ I_{\sigma} (t) \right]^2= R_0 \left[e\frac{dn_{\sigma}(t)}{dt} \right]^2.
\end{equation}
We now turn  to calculate the flux of particles per unit time with spin $\sigma$, $dn_{\sigma}(t)/dt $ at the first order in  $\dot{V}_{\sigma}(t)$.  From Eq. (\ref{eq:nSigmaAdiabatico}), we see that this is directly related to the fluctuation of the frozen occupation, $\delta \nfsig (t)$, under a small variation of the gates taking place in a small time interval $\delta t$, 
$\delta V_{\sigma} (t)= V_{\sigma}(t+ \delta t)- V_{\sigma}(t)$. Hence, 
\begin{equation}\label{nfsig}
\frac{d\nfsig(t)}{dt} = \lim_{\delta t \rightarrow 0} \sum_{\sigma} \frac{\delta n_{f \sigma}(t)}{\delta V_{ \sigma^{\prime}}(t)}
 \frac{\delta{V}_{\sigma^{\prime}}(t)}{\delta t}.
\end{equation}
Introducing the definition of the static frozen susceptibility.
\begin{equation}\label{defchi}
\chi^{\sigma \sigma^{\prime}}_t(0)= \frac{\delta n_{f \sigma}(t)}{\delta V_{ \sigma^{\prime}}(t)},
\end{equation}
Eq. (\ref{nfsig}) can be expressed as follows
\begin{equation}\label{nfsig1}
\frac{d\nfsig(t)}{dt} = \sum_{\sigma} \chi^{\sigma \sigma^{\prime}}_t(0) \dot{V}_{\sigma^{\prime}}(t).
\end{equation}

Therefore, if we keep only terms up to second order in $\dot{V}_{\sigma}(t)$ in Eq. \eqref{pjoulesigma}, we get
\begin{equation}
P_{\rm diss, \sigma} (t) = P_{\rm Joule, \sigma} (t)= R_0 \left[e\frac{dn_{f\sigma}(t)}{dt} \right]^2.
\end{equation}

Using Eqs. (\ref{1}) and (\ref{2})
and collecting the coefficients proportional to the different combinations of $\dot{V}_{\sigma}(t) \dot{V}_{\sigma^{\prime}} (t)$, we find that the following identities should be satisfied
\begin{eqnarray}
\label{flu-dis}
\frac{1}{2 h} \left\lbrace
\left[\chi^{\sigma \sigma}_t (0) \right]^2 +
\left[\chi^{ \overline{\sigma} \sigma}_t(0) \right]^2 
\right\rbrace &=& 
\Lambda^{\sigma \sigma} (t), \nonumber  \\
\frac{1}{ h} \left[\chi^{\uparrow \uparrow}_t(0) \chi^{\uparrow \downarrow }_t(0) +\chi^{\downarrow \downarrow}_t(0) \chi^{\downarrow \uparrow}_t(0)\right] &=&  \Lambda^{\uparrow \downarrow} (t)+ \Lambda^{ \downarrow \uparrow} (t).
\end{eqnarray}
The first of these identities has been discussed in Refs.  \onlinecite{re1,adia-nice,entro} for a non-interacting quantum dot, in which case, $\chi^{ \overline{\sigma} \sigma}_t(0)
=  \Lambda^{ \overline{\sigma} \sigma}(t)=0$. In the interacting case, this identity was originally derived by Shiba on the basis of Fermi liquid  theory in Ref. \onlinecite{ks}, for a system without magnetic field. This was 
later
generalized in Ref. \onlinecite{mich2} for a system with magnetic field.
The second identity is identically zero in the non-interacting case and, to the best of our knowledge, it has not been previously discussed in the literature. It is highly non-trivial and 
are a consequence of the fact that in the present model the spin fluctuations do not induce extra mechanisms of dissipation to the Joule law expressed by Eq. (\ref{pjoulesigma}).

\subsubsection{Generalized Korringa-Shiba relations}
These relations are basically combinations of the relations expressed by Eq. (\ref{flu-dis}). In fact, substituting the fluctuation-dissipation relations presented in Eqs. (\ref{flu-dis}) into the definitions of Eq. (\ref{lamc}), the following relations can be proved
\begin{eqnarray}
\label{eq:KScarga}
 \Lambda_c(t) & = &  \frac{h}{2} \left[
 \left( \chi^{\downarrow\downarrow}_t (0)+ \chi^{\downarrow\uparrow}_t (0) \right)^2 + 
 \left( \chi^{\uparrow\downarrow}_t(0) + \chi^{\uparrow\uparrow}_t (0) \right)^2
 \right],
\\
\label{eq:KSmag}
 \Lambda_m(t) & = &  \frac{h}{2} \left[
 \left( \chi^{\downarrow\downarrow}_t (0) - \chi^{\downarrow\uparrow}_t (0) \right)^2 + 
 \left( -\chi^{\uparrow\downarrow}_t(0) + \chi^{\uparrow\uparrow}_t (0) \right)^2
 \right],
\\
\label{eq:KScross}
 \Lambda_{cm}(t) &=& \frac{h}{2} \left[
 \left(\chi^{\uparrow\uparrow}_t (0) \right)^2 -
\left(\chi^{\downarrow\downarrow}_t(0) \right)^2 +
 \left( \chi^{\downarrow\uparrow}_t (0) - \chi^{\uparrow\downarrow}_t(0) \right) \right. \nonumber \\
& & \left.  \;\;\;\;  \times \left( \chi^{\downarrow\uparrow}_t(0) + \chi^{\uparrow\downarrow}_t(0) \right) \right].
\end{eqnarray}
Notice that Eq. (\ref{eq:KScarga}) is the Korringa-shiba (KS) law presented in Refs. \onlinecite{ks,mich2}. Here, we show that the assumption of a dissipation mechanism in the form
of the universal IJL in the driving problem with two parameters leads to the additional relations expressed in Eqs. (\ref{eq:KSmag}) and  (\ref{eq:KScross}).

\section{Results}

\subsection{Verifying the KS relations}
\begin{figure}[H]
  \includegraphics[width=\columnwidth]{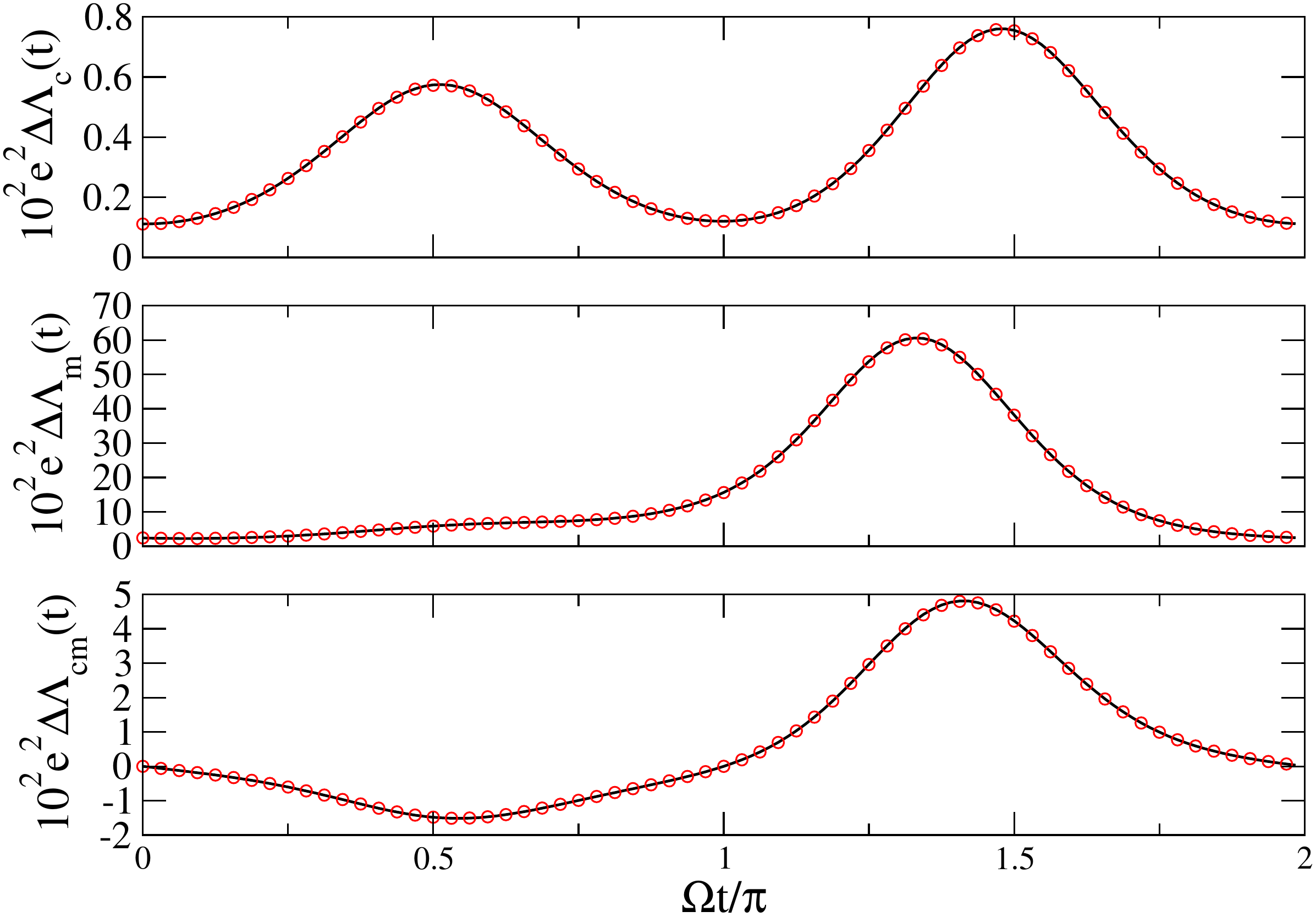}
  \caption{Generalized KS relations computed by NRG for $U=0.03$ and $\Delta=0.0008$. The driving protocol is  $V(t)=V_0 \sin(\Omega t), \; B(t)= B_0 \sin(\Omega t+\pi/4) + 0.0005$, with $V_0=0.006$, and $B_0=0.0003$.
 Energies are expressed in units of the bandwidth of the reservoir. 
  Top panel: Eq. \eqref{eq:KScarga}, middle panel: Eq. \eqref{eq:KSmag}, bottom panel: Eq.  \eqref{eq:KScross}. Solid lines correspond to the direct calculation of the Onsager coefficients 
  $\vec \Lambda_C $ and symbols to the evaluations from the frozen occupancies.}
  \label{fig:KS}
\end{figure}

The fluctuation dissipation relations, or, equivalently, the generalized KS relations introduced in the previous section have a very important outcome.
Namely, the dynamics of the driven quantum dot in contact to the reservoir at $T=0$ can be fully described by the knowledge of the frozen QD occupancy per spin  $n_{f,\sigma}(t)$.
In fact, given these occupancies, the static susceptibilities can be evaluated from Eq. (\ref{defchi}). Then, the fluctuation-dissipation or KS relations lead to  $\vec \Lambda (t)$ or
$\vec \Lambda_C (t)$. These coefficients enable the full characterization of the charge, spin and energy dynamics. Therefore, we start by verifying the fulfillment of the KS relations. 

In the non-interacting case with $U=0$, all local properties of the QD are characterized by the instantaneous spin-resolved  local density of states 
\begin{equation}\label{rho}
\rho_{\sigma}(\varepsilon, V_{\sigma})= \frac{2 \Delta}{\left(\varepsilon - V_{\sigma}(t) \right)^2+ \Delta^2 },
\end{equation}
where $\Delta$ depends on  the hybridization between the QD and the reservoir, assuming for the latter a constant density of states.
The frozen occupancy, the static susceptibility and the adiabatic Onsager coefficients can be easily calculated. The results are \cite{adia-nice,entro,rome}
\begin{eqnarray}\label{nonint}
n_{f,\sigma}(t) & = & \int \frac{d\varepsilon}{2\pi} f(\varepsilon)\rho_{\sigma}(\varepsilon, V_\sigma(t)), \nonumber \\
\chi_t^{\sigma \sigma^{\prime}} (0) &= & \delta_{\sigma, \sigma^{\prime}} \rho_{\sigma}(\mu), \;\;\;\;\;\;\;\;
\Lambda^{\sigma \sigma^{\prime}}(t) =  \delta_{\sigma, \sigma^{\prime}} \left[ \rho_{\sigma}(\mu) \right]^2,
\end{eqnarray}
where $f(\varepsilon)=\Theta(\mu-\varepsilon)$ is the Fermi-Dirac distribution function at $T=0$ and $\mu$ is the chemical potential of the reservoir.
From the previous expressions, we can readily verify that the KS relations are explicitly satisfied.

For the interacting case, we rely on numerical results obtained by numerical renormalization group (NRG). \cite{hewson,zitko} Results are shown in 
Figs. \ref{fig:KS},  where we see an excellent agreement between the direct calculations of the Onsager coefficients $\Lambda_c(t), \; \Lambda_m(t), \;  \Lambda_{c m}(t)$, and the calculation
of these coefficients from the evaluation of the static susceptibilities $\chi_t^{\sigma \sigma^{\prime}} (0)$ and the identities of Eqs. (\ref{eq:KScarga}),  (\ref{eq:KSmag}), and (\ref{eq:KScross}).

\subsection{Instantaneous occupancy}

\begin{figure}[H]
  \includegraphics[height=5cm]{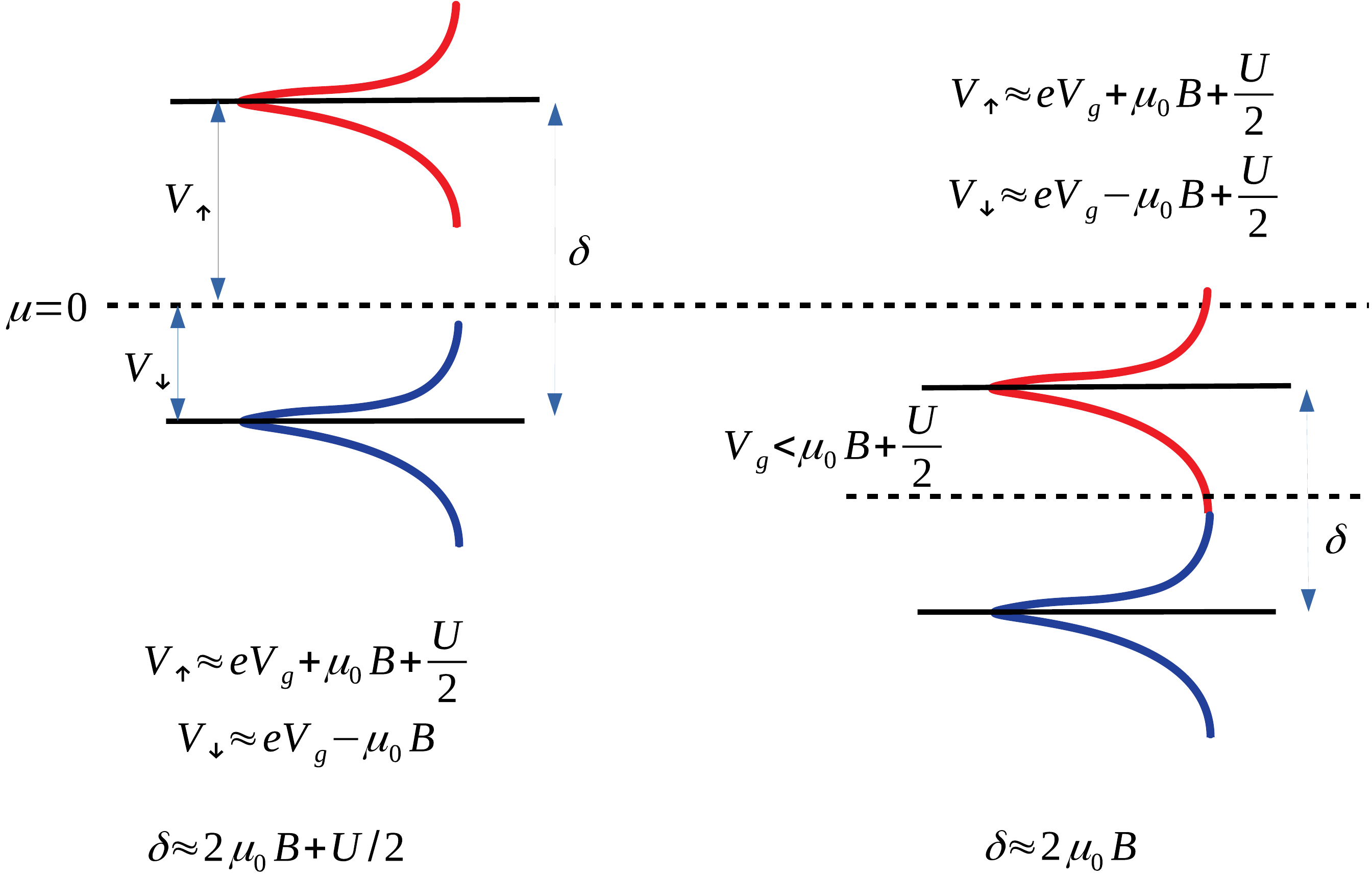}
  \caption{Sketch of the evolution of the occupancy of the QD. The energy of the singly and doubly occupied states  have an energy gap 
  $\delta$ which depends on the Coulomb interaction and on the magnetic field.}
  \label{fig:ocups_schema}
\end{figure}

The dynamics of the occupancy in the present system can be qualitatively understood in terms of a mean-field picture, where the Coulomb interaction term is replaced by $U\lbrace (n_{f,\downarrow} -1/2) \hat n_\uparrow + ( n_{f, \uparrow}-1/2) \hat n_\downarrow \rbrace$.  This leads to a local spin-resolved density of states obeying Eq. (\ref{rho}) upon the replacement $V_{\sigma} \rightarrow \tilde V_\sigma(t) = \varepsilon_0 + eV(t) + s_\sigma \mu_B B(t) + U( n_{f, \bar \sigma}-1/2)$, with the self-consistent relation
\begin{equation} \label{meanfield}
n^{MF}_{f, \sigma}(t)= \int \frac{d\varepsilon}{2\pi} f(\varepsilon)\rho_{\sigma}(\varepsilon,\tilde V_\sigma).
\end{equation}
It is important to notice that this quantity effectively depends on two parameters entering the definition of $\tilde{V}_{\sigma}$: $V_{\sigma}$ and $V_{\overline{\sigma}}$. This is in contrast to the non-interacting case
given by Eq. \eqref{nonint}, which depends only on $V_{\sigma}$.

A sketch of the evolution is indicated in Fig. \ref{fig:ocups_schema}, where the levels corresponding to the up and down occupancies are represented in red and blue, respectively. For fixed $B$, assuming a configuration
where the dot is initially singly occupied as in the left panel of the Fig., there is an energy gap between the occupied state with a given spin orientation (down spin in this configuration) and the state with the opposite spin orientation. Such energy gap depends, not only on the magnitude
of the Zeeman splitting (equal to $2 \mu_B B$), but also on the Coulomb energy (equal to $U  n_{f, \bar \sigma}$, in the mean-field description). The effect of changing the gate voltage is to move these two levels upwards or downwards
rigidly in energy, until a change in the occupation takes place. The sketch shown in the right panel of Fig.  \ref{fig:ocups_schema} corresponds to a protocol where $V_g(t)$ decreases, lowering the energy of the levels and enabling
the doubly occupancy. Changing in time the magnetic field implies a change in time of the Zeeman splitting.

\begin{figure}[H]
  \includegraphics[width=\columnwidth]{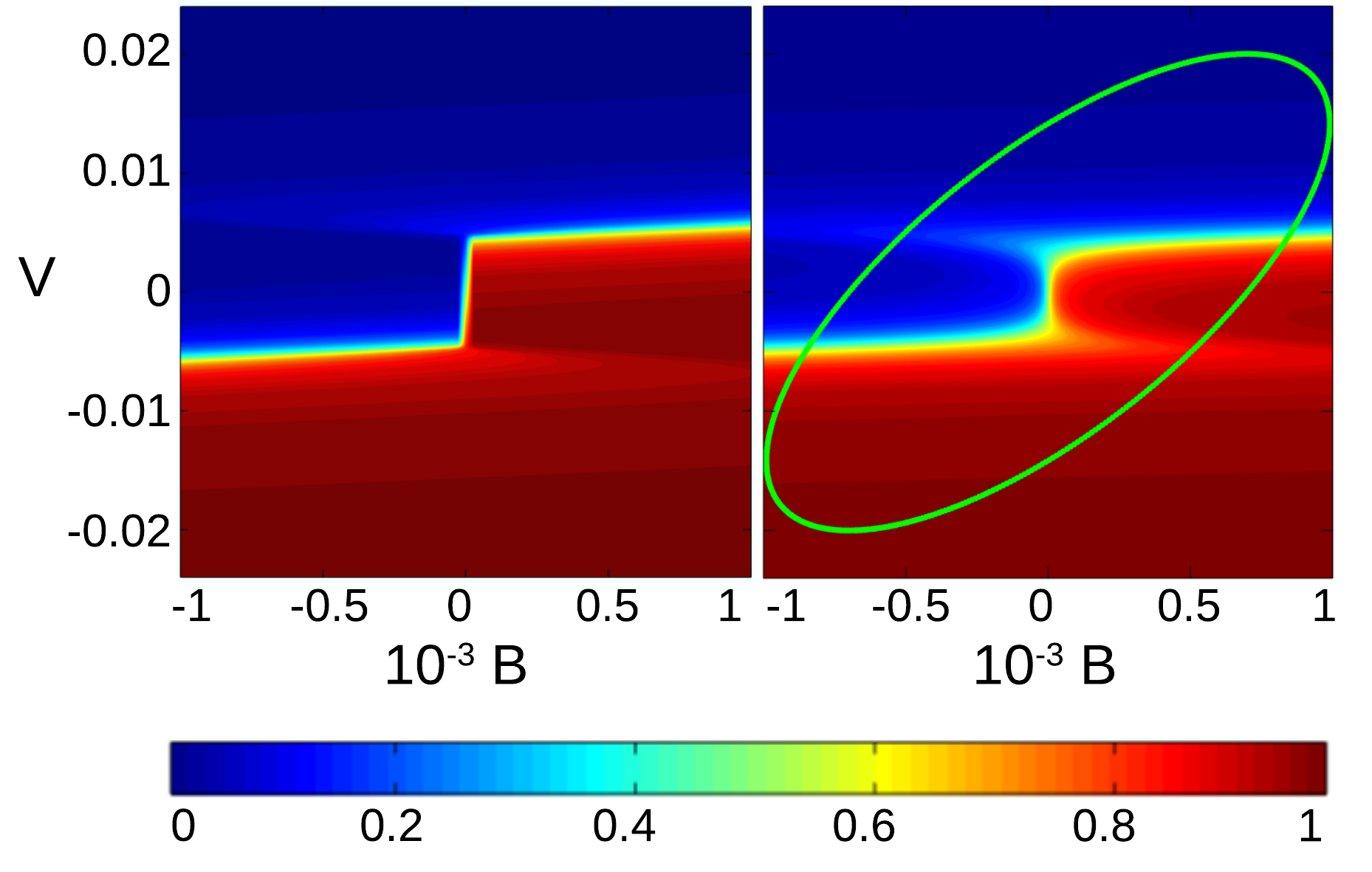}
  \caption{Occupancy map of $n_{f,\downarrow}$ in the mean field approximation for $U=0.01$ and $\Delta=0.0008$. The green elipse indicates the driving protocol  $V(t)=V_0 \sin(\Omega t), \; B(t)= B_0 \sin(\Omega t+\pi/4) + 0.0005$, with $V_0=0.02$, and $B_0=0.001$. Left and right panels correspond to results computed with  mean-field and NRG, respectively.
 Energies are expressed in units of the bandwidth $D$ of the reservoir. }
  \label{fig:ocdw}
\end{figure}

Results for the occupancy with down spin orientation, $n_{f, \downarrow}$ as function of the frozen parameters $V_g(t)$ and $B(t)$ are shown in Fig. \ref{fig:ocdw}, where the results obtained within the mean field description are compared with the ones
calculated numerically with the NRG method of Refs. \onlinecite{zitko,hewson}, considering a reservoir with a constant density of states and a bandwidth $D$ Notice that the occupancy with the opposite spin orientation is given by $n_{f,\uparrow}(V_g,B)=n_{f,\downarrow}(V_g,-B)$. We see that the 
mean-field approximation  is in very good qualitative agreement with the exact numerical calculation, which means that the exact results may be interpreted in terms of simple pictures of two moving levels as the ones sketched in
Fig. \ref{fig:ocups_schema}.

\begin{figure}[H]
  \includegraphics[height=5cm]{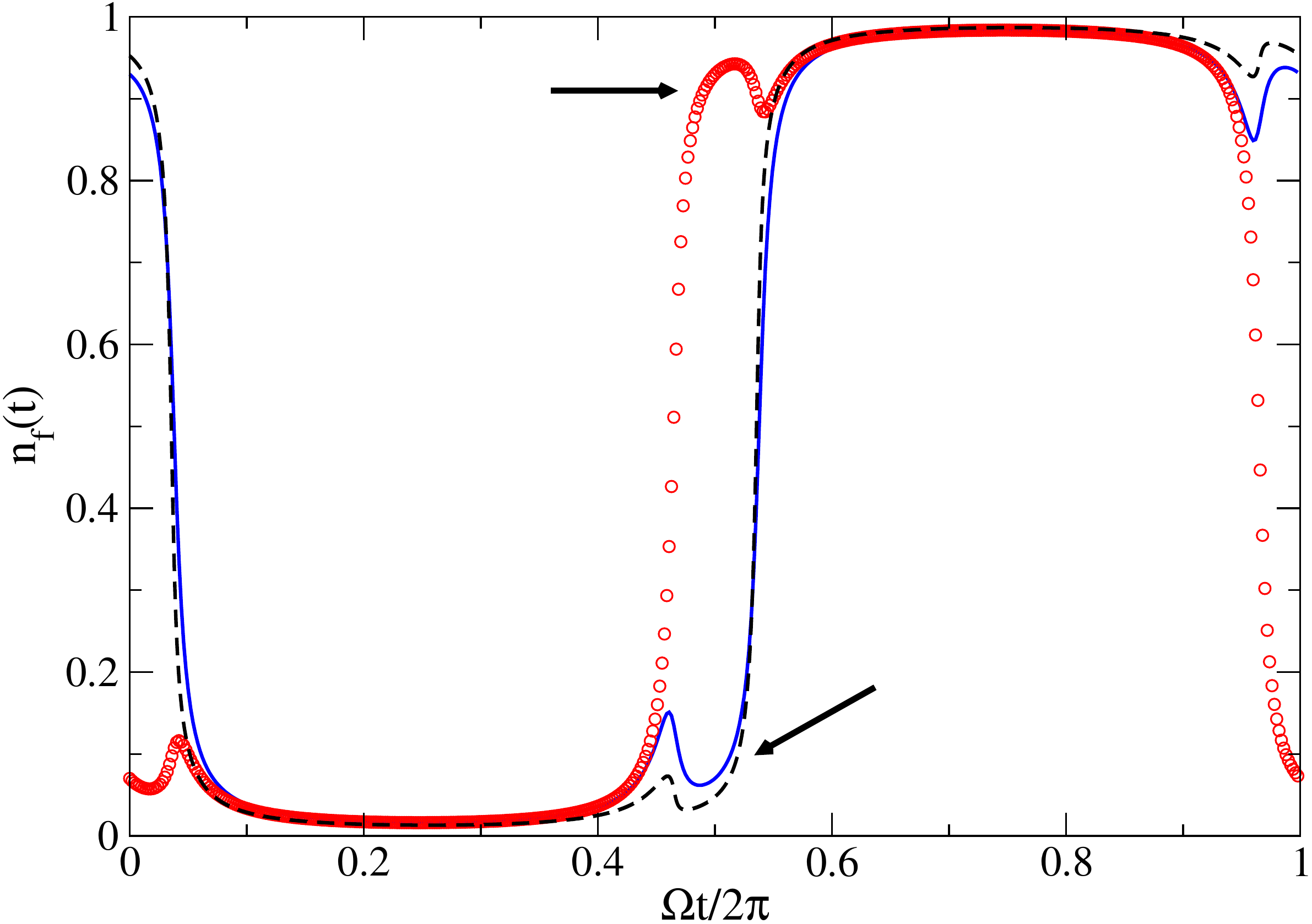}
  \caption{Frozen occupation along the curve of Figure \ref{fig:ocdw}. Solid line (blue): $n_{f\downarrow}(t)$ computed by NRG. Dashed line (black): $n_{f\downarrow}(t)$ computed in mean field approximation for parameters along the green curve of Fig. 4. Circles (red): $n_{f\uparrow}(t)$ computed by NRG.}
  \label{fig:occ_greencurve}
\end{figure}

The up and down frozen occupancies along the green curve in the previous figure are presented in Fig. \ref{fig:occ_greencurve}. Here, we also notice the good agreement between the NRG and mean-field description. This is because for the parameters chosen, the amplitude of the Zeeman splitting $2 \mu B$ is larger than the Kondo energy 
$k_B T_K= \sqrt{w_c^2 \pi U/2D^2} \exp\left(- \pi^2 w_c^2 U /8 D \right)$. \cite{zitko,hewson} Under these conditions, the Kondo effect is not 
robust, and the physics is dominated by Coulomb blockade in combination with the magnetic field, as described by the sketch of Fig. \ref{fig:ocups_schema}. This regime can be properly described by a simple mean-field approximation.

We
 can appreciate an interesting mechanism, which takes place just at the moment where the
occupancy change takes place (see arrows in the Fig.). This consists in a counter fluctuation in one of the occupancies when the other one changes from a filled to an empty configuration. This effect is a consequence of 
the shift in energy $U n_{f, \bar \sigma}(t) $ experienced by the electrons with spin $\sigma$, as $n_{f, \bar \sigma}(t) $ changes, which induces a  concomitant change in $n_{f, \sigma}(t) $. A similar effect was found for driving only 
in $V_g(t)$ and constant magnetic field (see Ref. \onlinecite{rome}). 

\subsection{Instantaneous Joule law and work exchange}
The fact that the KS relations are satisfied implies that the rate at which the total energy is dissipated follows an IJL with the universal resistance $R_0$. This is, precisely, expressed in Eq. (\ref{pjoulesigma}).

This does not necessarily mean that the total power developed in each spin channel follows the Joule law (\ref{pjoulesigma}).  
However, it can be directly verified from Eqs. (\ref{rho}) and (\ref{nonint}) that each spin component of the non-conservative power is purely dissipative, and follows the instantaneous Joule law per spin channel in the non-interacting limit ($U=0$):
\begin{equation}
 P_{{\rm non-cons},\sigma} (t)= P_{\rm diss, \sigma} (t) = P_{\rm Joule, \sigma} (t), \;\;\;\;\;\;\; U=0.
\end{equation}

\begin{figure}[H]
  \includegraphics[width=\columnwidth]{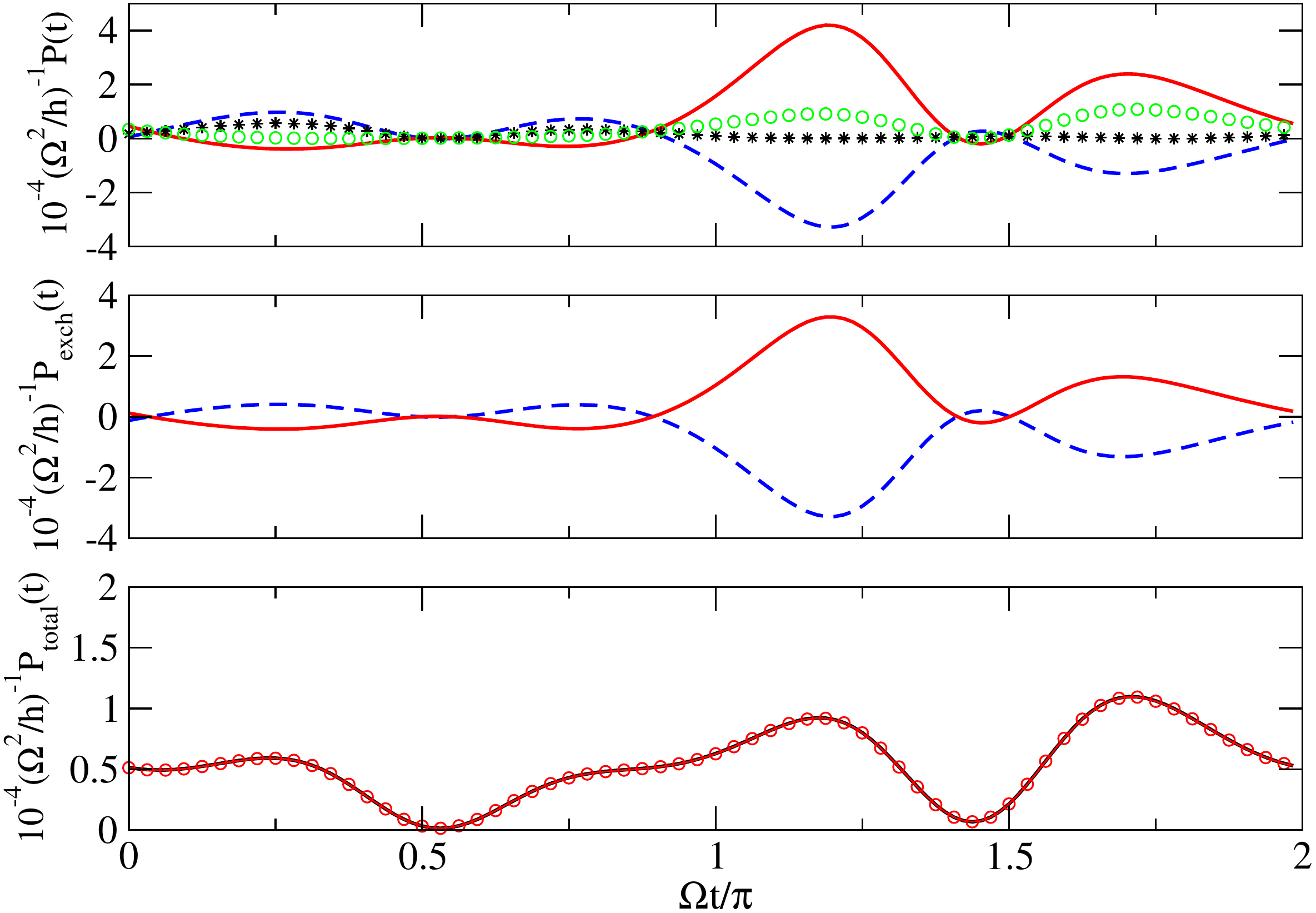}
  \caption{Computed powers with NRG. $U=0.03$ and $\Delta=0.0008$, under the driving protocol $V_g(t)=V_0 \sin(\Omega t)$, $B(t)=B_0 \sin(\Omega t+\pi/4) + 0.0005)$ with
  $V_0=0.006$ and $B_0=0.0003$. Energies are expressed in units of the bandwidth of the  reservoir.  Top panel: plots in solid red, dashed blue, green circles and black stars correspond, respectively,
  to  $P_\uparrow(t)$, $P_\downarrow(t)$,  $P_{\rm Joule,\uparrow}(t)$, $P_{\rm Joule,\downarrow}(t)$.
 Middle panel:
  Plots in solid line red and dashed blue correspond to  $P_\uparrow(t) - P_{Joule,\uparrow}(t)$ and $P_\downarrow(t) - P_{Joule,\downarrow}(t)$, respectively.
Bottom panel: the total power $P_\uparrow(t) + P_\downarrow(t)$ is plotted in solid black and coincides with the total total Joule dissipation. $P_{\rm Joule,\uparrow}(t) + P_{\rm Joule,\downarrow}(t)$, which is
plotted in red circles.  }
  \label{fig:potencias}
\end{figure}

For the interacting system ($U\neq 0$) the fulfillment of the KS relations discussed previously, indicate that the dissipative component of the non-conservative power per spin channel follows a IJL,
as expressed in Eq. (\ref{pjoulesigma}). However, in the interacting case, the non-conservative components may contain an extra term, denoted by $P_{\rm exch} (t)$ in Eq. (\ref{1}).  
Fig. \ref{fig:potencias} shows the behavior of the spin-resolved non-conservative power $P_{{\rm non-cons},\sigma} (t)$ in the interacting system with two-parameter driving. In the upper panel of the Fig., this power 
is shown, along with the dissipative component $P_{\rm Joule, \sigma} (t)$ for the two spin orientations. We can see that, although $P_{{\rm non-cons},\sigma} (t)$ differs from $P_{\rm Joule, \sigma} (t)$, the corresponding difference is, precisely, the exchanged power between the two spin orientations $P_{\rm exch} (t)$, which is shown in the middle panel. The lower panel
of the Fig. displays the total non-conservative power, where we see that it is fully dissipative and coincides with  $\sum_{\sigma} P_{\rm Joule, \sigma} (t)$.
Interestingly, we can see in Fig. \ref{fig:potencias} that $P_{\rm exch}(t)$ has a non-vanishing mean value when integrated over a period, as indicated in Eq. (\ref{netex}).

\begin{figure}[H]
  \includegraphics[height=6cm]{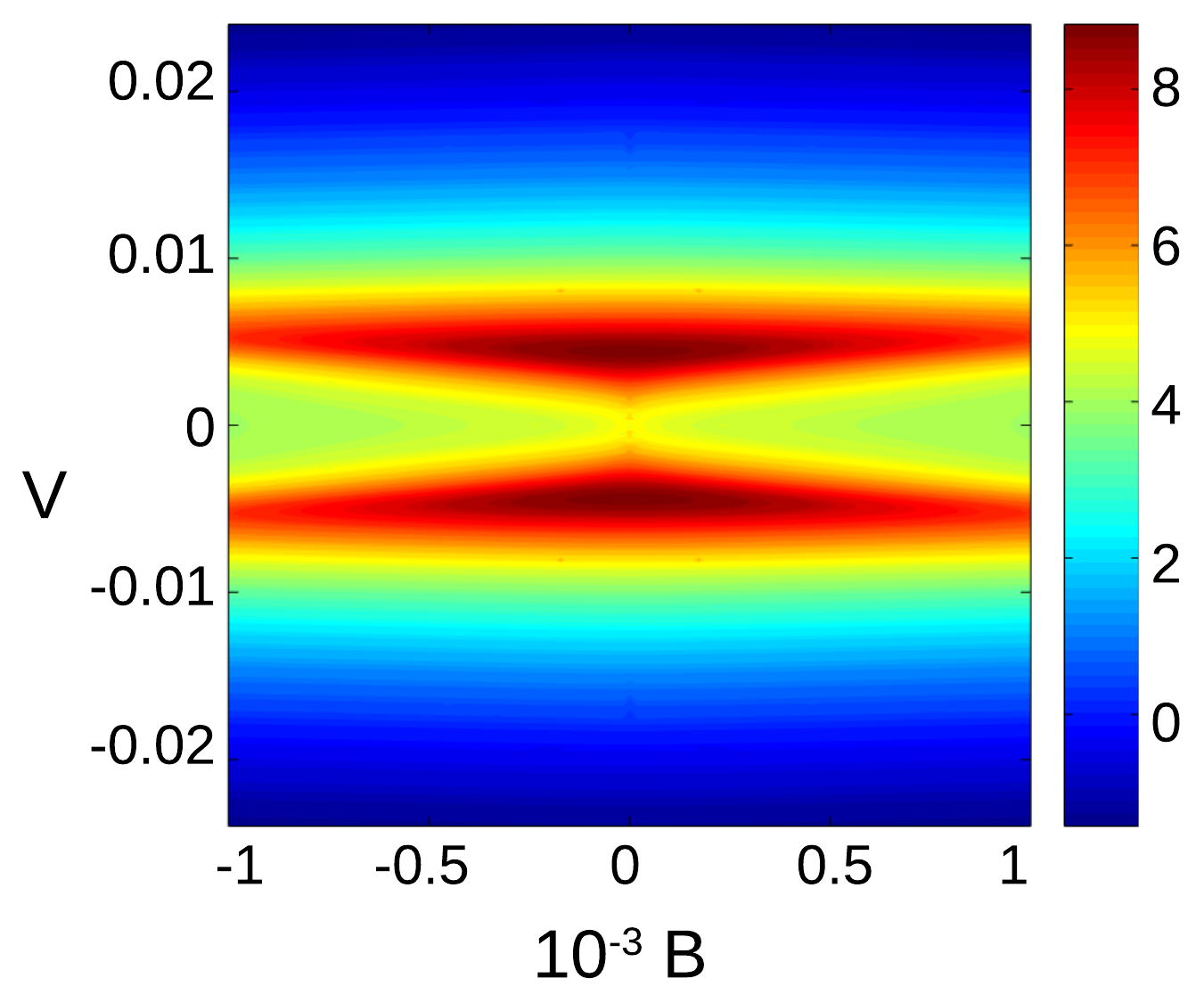}
  \caption{Minimal eigenvalue (in log scale) of the matrix $\vec \Lambda (\vec V) $ for $U=0.01$ and $\Delta=0.0008$. Energies are expressed in units of the bandwidth of the reservoir. }
  \label{fig:eigen}
\end{figure}
We now turn to analyze the structure of the matrix $\vec \Lambda (\vec V)$ for the interacting QD. As discussed in relation to Eq. \eqref{eq:lam}, all the eigenvalues of this matrix are positive in the 
dissipative system and  we can directly verify this property.
Fig. \ref{fig:eigen} shows the minimum eigenvalue $\vec \Lambda (\vec V)$.

  It is interesting to notice that the largest values of the minimal eigenvalue correspond to parameters favoring the charge fluctuation of the QD. Notice that the highest values are concentrated on values of $V_g(t)$ for which the two many-body levels of the quantum dot, which  separated in the energy $\delta$ indicated in Fig. \ref{fig:ocups_schema}, become aligned with the Fermi energy of the reservoir. For those parameters, the change in the occupation for small changes in $V_{\sigma}(t)$, thus the charge current $I_{\sigma}(t)$ achieves high values. Hence, the dissipation following the Joule law is also large.

\subsection{Non-equilibrium polarized charge population}
\begin{figure}[H]
  \includegraphics[height=5cm]{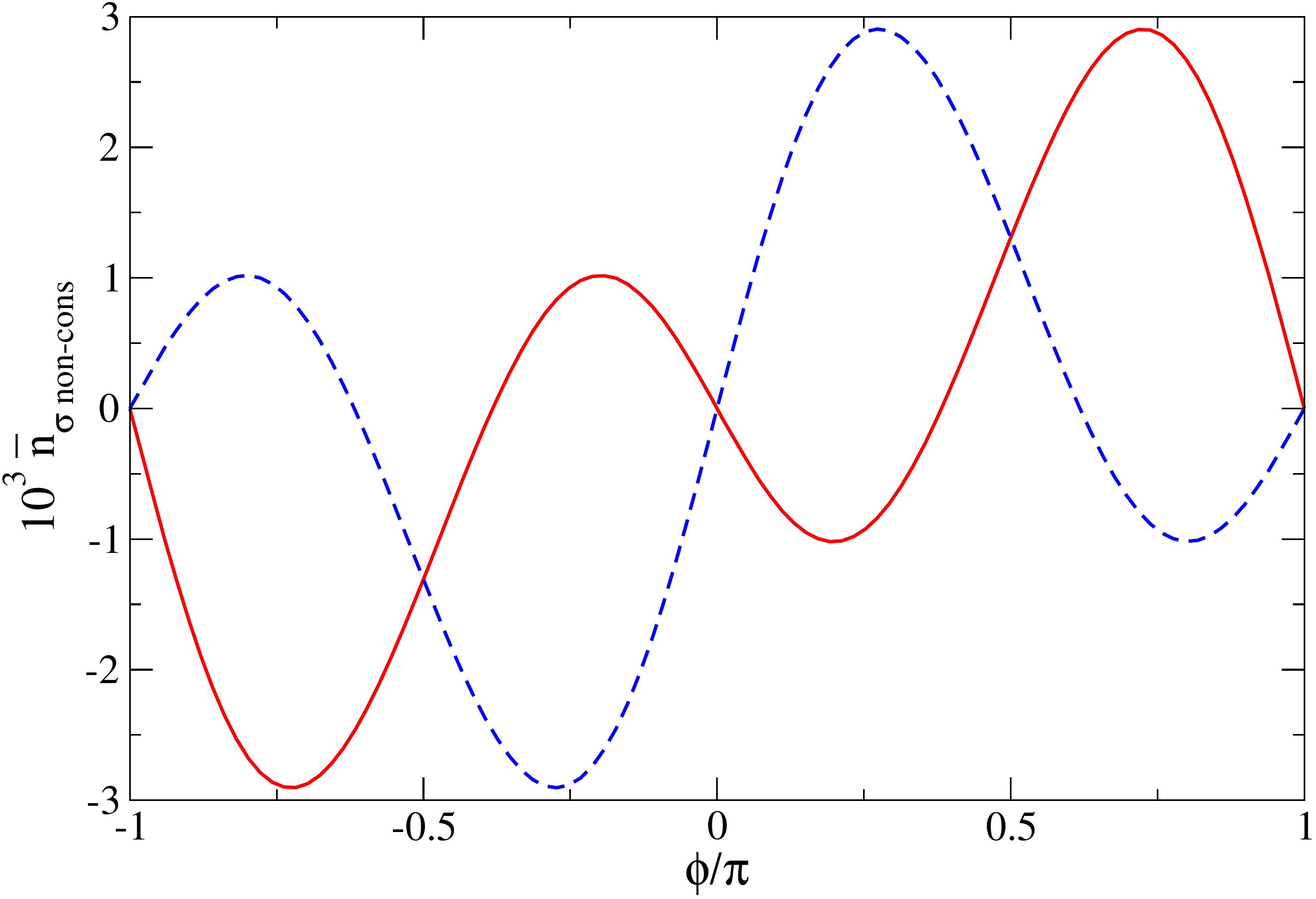}
  \caption{Mean charge accumulation for each spin orientation  for $U=0.03$ and $\Delta=0.0008$, computed with NRG, as a function of the phase difference $\phi$ of the driving protocol $V_g(t)=V_0 \sin(\Omega t), B(t)= B_0 \sin(\Omega t+\phi) + 0.0005)$,
 for $V_0=0.006$ and $B_0=0.0003$. Dashed line (blue): $\sigma = \downarrow$. Solid line (red): $\sigma = \uparrow$.}
  \label{fig:vsfase}
\end{figure}

In contrast to the case where the QD is driven by a single parameter, it is possible in the present case to induce a net non-equilibrium polarized charge in the quantum dot. Akin to the pumped charge between two 
reservoirs in quantum dots driven by two parameters, this quantity has a geometric character, as expressed by Eq. (\ref{non-cons}). Importantly, for the dynamics to depend actually on two parameters, it is necessary to have 
a many-body interaction in the quantum dot. In fact, notice that in the non-interacting case, Eqs. (\ref{rho}) and (\ref{nonint}) make it explicit that the dynamics of $n_{f,\sigma}(t)$  depends only on the single 
parameter $V_{\sigma}(t)$ and it is completely independent from $V_{\overline{\sigma}}(t)$. Instead, in the interacting case, even at the simple mean-field level it can be seen that the evolution of the occupation  $n_{f,\sigma}(t)$ depends on the two parameters  $V_{\sigma}(t)$ and $V_{\overline{\sigma}}(t)$ [see  Eq. (\ref{meanfield})].  Results for 
the net charge accumulation with each spin orientation, calculated with NRG, are shown in Fig. \ref{fig:vsfase}.  It is important to notice, that such a non-equilibrium charge 
accumulation takes place against the condition imposed by the chemical potential of the reservoir.
As stressed before, this quantity has a geometric nature, similar to
the pumped charge in two-parameter two-terminal driven systems. \cite{avron, brouwer} This also bears resemblance to the geometric magnetism discussed in Ref. \onlinecite{miche}
and it is interesting at this point to discuss similarities and differences between that work and the present contribution. Both works have a common point of view in the general approach, in the sense that in both cases, the adiabatic dynamics is addressed as a linear-response treatment in the velocities $\dot{\bf V}$.  In Ref. \onlinecite{miche} the name
"geometric magnetism" is used to characterize the antisymmetric component of Onsager matrix $\vec{\Lambda}$. This is because the resulting induced force in that case has the
formal structure of a Lorentz force, like the one experienced by a charged particle in a magnetic field. In such adiabatic dynamics the field is not a real, but a ficticios one.  In the system we analyze here, the adiabatic dynamics has a purely symmetric $\vec{\Lambda}$. Since the induced force is associated to a real spin polarization in the quantum dot, we have real magnetism in the present problem. In both cases, the induced net forces can be expressed in terms of a geometric quantity, like the contour integral of Eq. (\ref{non-cons}). The latter
is conceptually similar to the pumped charge considered in Refs. \onlinecite{avron, brouwer}. Importantly, the dynamics ruled by an antisymmetric Onsager matrix is non-dissipative,
while the one ruled by a symmetric $\vec{\Lambda}$ is dissipative.

\section{Summary and conclusions}
We have investigated the adiabatic dynamics of a QD with many-body interactions connected to a single reservoir at $T=0$ and under the effect of two-parameter driving with a gate voltage and a magnetic field.
This induces time-dependent flow of charge, spin and energy through the contact between the QD and the reservoir.  Under these conditions, the net dynamics is fully dissipative. \cite{dis}
We have verified that energy is instantaneously dissipated in the form of a Joule law with a
universal resistance $R_0$. This was previously discussed in the framework of non-interacting, \cite{re1,adia-nice,entro},
 as well as interacting systems under single-parameter driving.\cite{rome}

Here, we have derived fluctuation-dissipation relations for the adiabatic responses in the framework of Ref. \onlinecite{adia}, which constitute generalizations of the so-called Korringa-Shiba law, previously derived for Fermi liquids. \cite{ks,miche}
We showed that in the presence of many-body interactions, other interesting effects take place as a consequence of the two-parameter driving. These are the net work exchange between forces as well as the non-equilibrium polarized charge population with geometric nature, akin to quantum pumps. \cite{avron,brouwer}  These features are amenable to be explored experimentally in quantum capacitors like those studied in Refs. \onlinecite{qcap1,qcap2,qcap3,qcap4}. In fact, these systems are constructed in the quantum Hall regime,
under the effect of  a strong magnetic field. By introducing a time-dependent component in this magnetic field, along with the variation of the gate voltage would  lead to a scenario like the one studied in the present work.  At finite $T$, the effects we have studied could be relevant in the implementation of driving protocols for thermal machines, 
\cite{mach1,mach2,mach3,mach4,mach5,mach6,mach7} as well as in the discussion of shortcuts to adiabaticity.\cite{shcut1,shcut2,shcut3}

\section*{Acknowledgments}
We acknowledge support from CONICET, Argentina  and  the Alexander von Humboldt
Foundation, Germany (LA). We are sponsored by  PIP-RD 20141216-4905 
of CONICET  and PICT-2014-2049 (LA and PTA).

\appendix
\section{Adiabatic dynamics}
At first order in the driving parameters, the expectation value of an operator $\hat A(t)$ is given by [see Eq. 5 of Ref. \onlinecite{adia}].
\begin{equation}
    \expval{\hat A(t)} = \expval{\hat A}_t - \intKubo \expval{
    \comm {\hat A(t)} {\hat{\bf{F}}(t')} }_t \bf{\dot V}(t),
\end{equation}
where $\expval{A}_t$ means the expectation value of $\hat A$ computed with the equilibrium density matrix of the frozen Hamiltonian $\H(t)$, and $\hat{\bf F}=-\partial \hat{H}/\partial{\bf V}$.

Applying this theory to the $\hat n_\sigma(t)$ operators with the dot Hamiltonian $\H_{dot}=\sum_\sigma (\varepsilon_0+V_\sigma(t))\hat n_\sigma + U(\hat \nup-1/2)(\hat\ndw-1/2)$ (with $\lbrace s_\uparrow, s_\downarrow = 1, -1 \rbrace$) we define the values
\begin{eqnarray}
\Lambda^{\sigma \sigma'}(t)
&=& -\intKubo \expval{\comm{\hat n_\sigma(t)}{\hat n_{\sigma'}(t')}} \nonumber \\
&=& -i \intInf d\tau \tau \Theta(\tau) \expval{\comm{\hat n_\sigma(\tau)}{\hat n_{\sigma'}(0)}}
\end{eqnarray}

And so we have that
\begin{equation}
    n_{\sigma}(t)=n_{f\sigma}(t) + e\sum_{\sigma'} \Lps(t) V_{\sigma'}(t)
\end{equation}

with the notation $n_{f\sigma}(t) = \expval{\hat n_\sigma}_t$.


\end{document}